# A quality assurance framework for real-time monitoring of deep learning segmentation models in radiotherapy

**Authors:**

Xiyao Jin[1]

Yao Hao[1]

Jessica Hilliard[1]

Zhehao Zhang[1]

Maria A. Thomas[1]

Hua Li[1]

Abhinav K. Jha[2,3]

Geoffrey D. Hugo[1]

**Affiliations:**

1. Department of Radiation Oncology, Washington University in St. Louis School of Medicine, St. Louis, Missouri
2. Department of Biomedical Engineering, Washington University in St. Louis, St. Louis, Missouri
3. Mallinckrodt Institute of Radiology, Washington University in St. Louis, St. Louis, Missouri

**Corresponding author:**

Geoffrey D. Hugo, PhD

Department of Radiation Oncology

Washington University School of Medicine

314-362-2639

gdhugo@wustl.edu

# Abstract

### Purpose

To develop a quality assurance (QA) framework that can be deployed for routine or continuous monitoring of input-domain shift and the performance of automatic segmentation algorithms without manual contours in radiotherapy.

### Methods

In this work, cardiac substructure segmentation was used as an example task to establish a QA framework. A benchmark dataset consisting of Computed Tomography (CT) images along with manual cardiac delineations of 241 patients undergoing radiotherapy were collected, including one "common" image domain and five "uncommon" domains. Segmentation models were tested on the benchmark dataset for an initial evaluation of model capacity and limitations. An image domain shift detector was developed by utilizing a trained Denoising autoencoder (DAE) and two hand-engineered features. Another Variational Autoencoder (VAE) was also trained to estimate the shape quality of the auto-segmentation results. Using the extracted features from the image/segmentation pair as inputs, a regression model was trained to predict the per-patient segmentation accuracy, measured by Dice coefficient similarity (DSC). Different regression techniques were investigated including naïve linear regression, Bagging, Gaussian process regression, random forest, and gradient boost regression. Out of the 241 patients, 60 were used to train the segmentation models. The rest of patient cases were split into 120 and 61 for the training and testing of the regression models. The framework was tested across 19 segmentation models, including 18 CNN-based and one transformer-based model to evaluate the generalizability of the entire framework.

### Results

When using bootstrap aggregating as linear regression models, the predicted DSC of regression models achieved a mean absolute error (MAE) ranging from $0.036$ to $0.046$ across 19 tested segmentation models, with an averaged MAE of $0.041 \pm 0.002$, which was the highest among other investigated regression techniques.

When tested on the benchmark dataset, the performances of all segmentation models were not significantly affected by scanning parameters: FOV, slice thickness and reconstructions kernels. When input images were added with Poisson noise, CNN-based segmentation models demonstrated a decreased DSC ranging from 0.07 to 0.41, while the transformer-based model was not significantly affected (ANOVA, $\mathrm{p} = 0.99$).

### Conclusions:

A quality assurance framework for monitoring cardiac substructure segmentation models was proposed, including three components: 1) a comprehensive benchmark dataset to evaluate the model robustness and limitation. 2) DAE based techniques and hand-engineered features for image-domain shift quantification and segmentation-shape assessment. 3) a regression model that, when input the extracted features, makes clinically acceptable predictions of segmentation performances.

# I. Introduction

In recent years, deep-learning (DL) based image analysis methods have been extensively studied and are setting the state-of-the-art on a broad range of medical image analysis. However, training a DL model usually needs a large dataset to achieve good performance. It is difficult and expensive to acquire (labeled) clinical data due to privacy restrictions or large amounts of labor to extract and label the data. Most researchers are therefore constrained to rely on public datasets, which are typically research cohorts that differ from a clinical setting[1]. Public datasets usually contain images acquired under a controlled research protocol. However, in a clinical setting, even for a single task and body site, image appearance can vary widely due to factors such as different imaging protocols, reconstruction algorithms, abnormal anatomies, and artifacts from implanted devices. Imaging protocols may change or vary based on clinical needs. Furthermore, entire scanners may change due to replacement.

While deep neural networks have shown strong performance in learning input-output mappings within the expected distribution of the training data, they demonstrate remarkable performance degradation when the domain of the test dataset deviates from that of the training set (i.e., data domain shift).[1–6] Several attempts have been made to explicitly investigate the robustness of data domain shifts in a DL model. For example, Martensson et al.[1] investigated the performance of a DL model for visual ratings of atrophy in brain MRI data from five different cohorts. They found that model trained on research data with harmonized protocols performed well but failed in other data, and training with more heterogeneous training data improved the model robustness. Zech et al.[7] found that a CNN trained for pneumonia screening on chest X-rays generalized poorly to new cohorts. In a study of limited angle tomography, Huang et al.[8] investigated the robustness of a DL model to input perturbations like Poisson noise and adversarial attack and found that adding Poisson noise to the training data improved the model robustness to Poisson perturbation but had minimal improvement on adversarial attack. [910]

These prior works suggest that for safe deployment of DL algorithms in clinics, it is vital that we obtain a holistic evaluation of model robustness towards a broad range of source of variabilities in clinical images and build a quality assurance (QA) system that routinely monitors DL algorithms performance for potential changes in input data after deployment. Vandewinckele et.al.[11] emphasized on the importance of QA for machine learning based applications in radiotherapy and developed recommendations on the implementation of automatic segmentation. Their work found the importance of both case-specific QA and routine model QA, which detects outlier case (e.g. abnormal anatomy) and systematical image changes (e.g. scanner parameter changes), respectively. Several attempts have been made to evaluate the risk in the clinical adoption of automated contouring using failure mode and effects analysis[12,13]. These and similar guidelines[14] recommend evaluation of performance by comparison to expert contours in a benchmark dataset for initial commissioning and visual inspection case by case for ongoing QA. This is a labor-intensive process which focuses on consistency of the algorithm performance against the benchmark and makes it challenging to routinely assess performance relative to changes in the imaging system. There have been studies outside of radiation therapy exploring the techniques of building QA systems using public datasets[15–18]. Other guidelines have advocated for detection of data drift[19]. Soin et.al. [20] utilized variational autoencoder (VAE) and image metadata to detect data drift in the task of chest x-ray image classification. Rabanser et.al. [21] investigated the techniques of dimension reductions and hypothesis testing to quantify the magnitude input data shift. Liu et.al.[22] utilized VAE to capture the shape statistics of "good" segmentation, which raises alarms when low quality segmentation occurs due to large reconstruction errors. Robinson et.al.[23] demonstrated that reverse classification accuracy has the potential for fully automatic segmentation QA.

The goal of this work is to develop a QA approach for monitoring the performance of the DL-based auto-segmentation that can be deployed for routine or continuous monitoring without needing to generate expert contours on new images at each monitoring point. We used cardiac substructure segmentation in breast cancer radiotherapy as a well-understood example problem to establish a framework of robustness and quality assurance. The key idea of this work is firstly using a comprehensive benchmark dataset to "probe" the limitation of a given segmentation model, thus evaluating its robustness against different types of image domain shifts or image quality degeneration. Then the evaluation results are implicitly "memorized" by a regression model, which is utilized to predict the performance of the segmentation model given new clinical input data. The key contributions of this work include: (1) a benchmark test dataset that comprises a clinically realistic source of domain variations to simulate the clinic setting. (2) a clinically feasible QA algorithm that incorporates both model robustness evaluation and automatic model performance monitoring. (3) The proposed QA algorithm generates per-patient predictions, providing the functionality to perform both anomaly detection and data-drift detection. (4) The proposed QA algorithm treats the auto-segmentation model as a black-box, without the need to access the internal working principles of the auto-segmentation algorithm.

## II. Methods

The overall pipeline of the QA system is shown in Figure 1. The QA system consists of 3 components: benchmarking of segmentation accuracy, a predicted shape quality assessment, and an image domain shift detector. The outputs of these three components are then used to build a regression model, denoted by G, with the goal of predicting the performance of the deployed segmentation model. Benchmarking of segmentation accuracy is estimated using a clinically realistic, labeled benchmark dataset. The segmentation outputs $\mathrm{S_{pred}}$ will be compared with ground-truth labels $\mathrm{S_{true}}$ to calculate the segmentation accuracy $\mathrm{Y_{true}}$. Separately, a set of CT images, denoted by the set $I$, are input through an image domain shift detector that comprises a pre-trained DAE and two hand-engineered features to quantify information regarding image appearance shift, intensity shift, and image noise level, resulting in three scalar features $X_{Shift}$, $X_{Intensity}$ and $X_{Noise}$, respectively. The segmentation output $S_{pred}$ is separately input through a shape quality estimator, which utilizes another pretrained VAE to detect out-of-distribution shapes, yielding a scalar feature $X_{Shape}$. The four feature outputs from the image domain shift detector and shape quality estimator are used as the inputs to train a regression model G to predict the segmentation accuracy $Y_{true}$.

At the inference stage, new clinical data $I_{clinic}$, along with its segmentation result $S_{pred}$, are input through the entire QA pipeline. Eventually a predicted segmentation accuracy $Y_{pred}$ will be generated to predict the model performance.

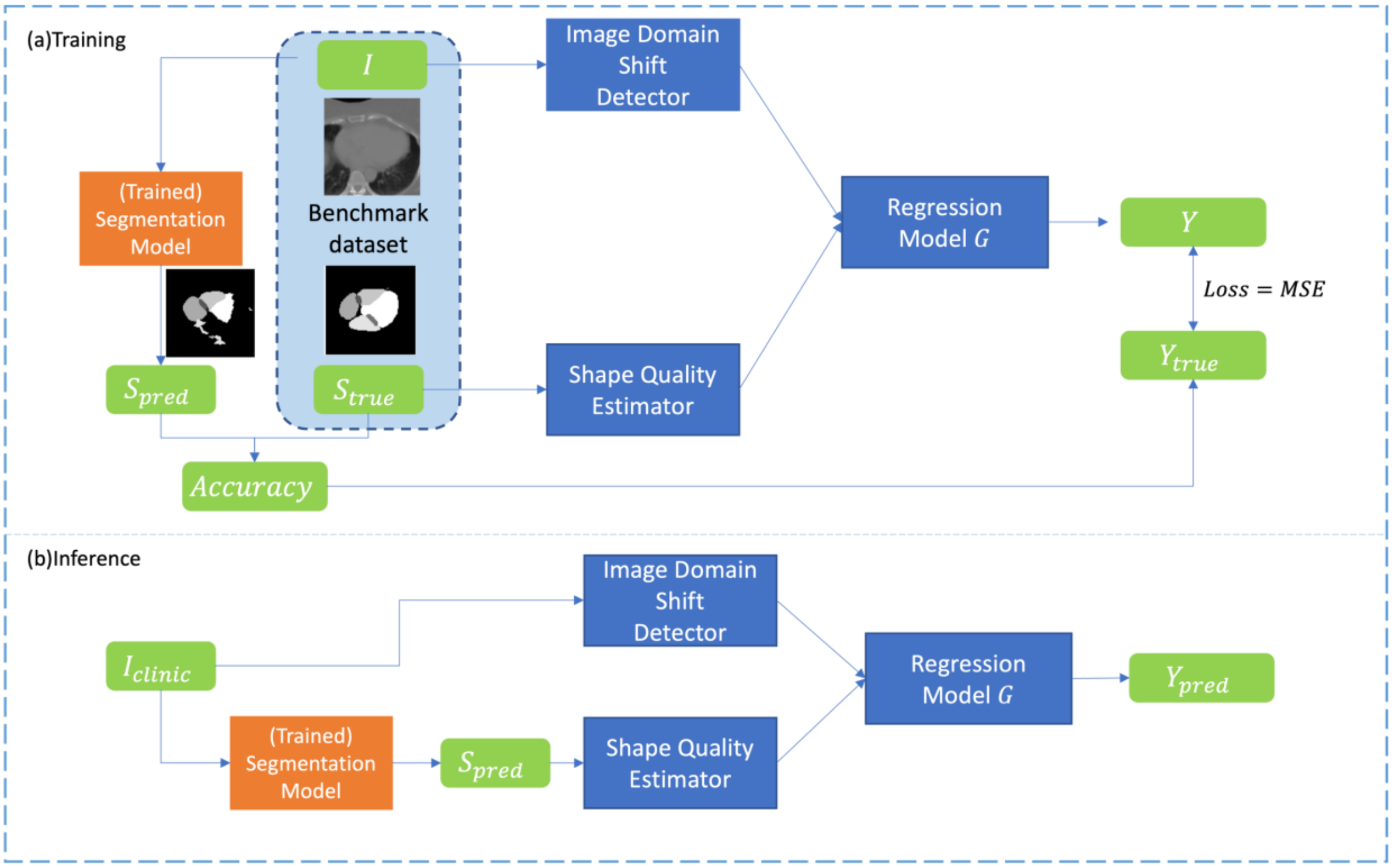

*Figure 1. The overall pipeline of the proposed QA system. The goal is QA of the deployed segmentation system (orange box). (a) At the QA algorithm training stage, the collected benchmark dataset containing CT images $I$ and ground-truth labels $S_{true}$ were separately used to train an image domain shift domain detector and a shape quality estimator, the output of which was used as inputs to train a regression model. (b) At the inference stage, the system takes new clinical data $I_{clinic}$ as input and output a predicted segmentation accuracy.*

## A. Benchmark Dataset

In this study, to evaluate the performance and limitations of a segmentation algorithm for a specific radiotherapy application, we designed and collected a benchmark dataset which contains chest CT images from various domains.

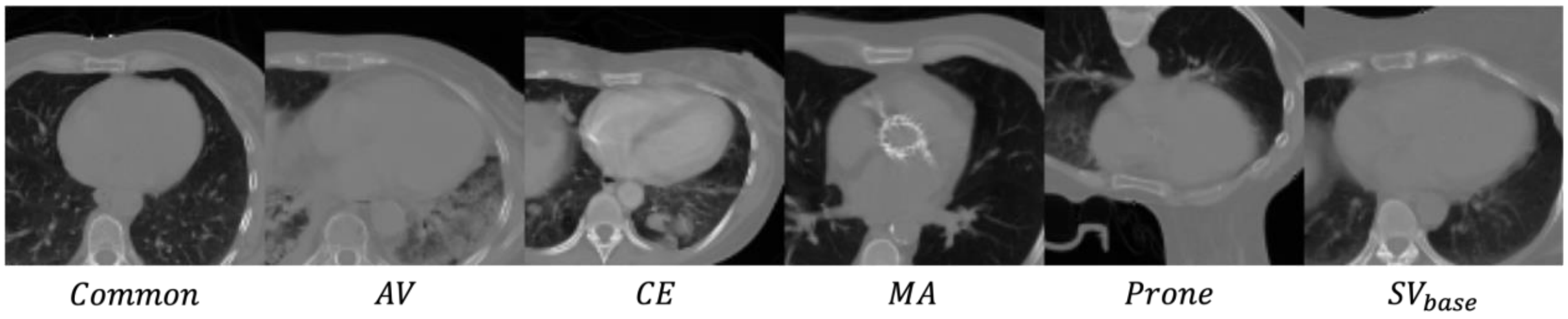

*Figure 2. Example images from the different domains of the benchmark dataset. The six domains are: (1) Common, where images were acquired using routine scanner protocols and without anatomical abnormal. (2) AV, where images contain anatomical variation or external implants. (3) CE, which contains contrast CT images. (4) MA, where images contain metal artifact. (5) Prone, where patients were in prone position when scanning and were flipped in image preprocessing. (6) SV, where the raw imaging data of 20 patients were collected to reconstruct CT images with various scanning parameters.* $SV_{base}$ *was reconstructed using the same scanner setting as the " common" domain.*

A total of 181 chest CT images of patients undergoing breast cancer radiotherapy were collected from our clinical database. The benchmark dataset consisted of one $common$ domain, and multiple $uncommon$ domains. The examples images from each domain are shown in Figure 2. The $common$ domain contains 39 non-contrast CT images with no obvious visual abnormal anatomy, no major metal implant, supine position, and with a field of view (FOV) of 600mm*600mm, slice thickness of 3mm and a reconstruction kernel of Siemens Qr43. We collected the dataset in a manner that tries to constrain each $uncommon$ domain with only one type of variation compared to the $common$ domain. The $uncommon$ domains were further categorized into the following five domains:

1) **Anatomical Variation** (AV): This domain contains 74 patients with either pathological variation (e.g., hemidiaphragm elevation) or variations due to external sources (non-metal implant devices, surgery, such as mastectomy). These variations were classified by an expert observer.
2) **Contrast Enhanced** (CE): In most cases intensity deviation in CT images is caused by the injection of contrast agent. Here we collected 20 contrast CT images. The average Hounsfield Unit (HU) of cardiac area, masked by manual cardiac delineations, ranges from 87 to 202, with a mean HU of 139 across 20 patients.
3) **Metal Implants** (MIP): CT images of 12 patients with artificial implants that causes metal artifact, such as tissue expander and pacemaker or cardioverter defibrillator along with other high contrast artifact, such as a port, were collected.
4) **Prone Position** (Prone): CT images of 16 patients with prone positions were collected. These images have no other type of variations compared to the $\mathrm{common}$ domain, except patient position. During the preprocessing, these images were additionally flipped to simulate the "pseudo supine" position.
5) **Reconstruction-protocol Variation** (RPV): The raw imaging data of 20 patients were collected and reconstructed through the scanner's built-in software multiple times using different reconstruction parameters. We firstly reconstructed the raw data using the default settings the same as the $\mathrm{common}$ domain, generating the $SV_{base}$ domain. Each time, we changed one parameter relative to the common settings and applied them to all the 20 patients. Specifically, we investigated the FOV of 300mm*300mm, 400mm*400mm and 500mm*500mm, the slice thickness of 1.5mm, 2mm, 4mm and 5mm, and seven reconstruction kernels which are listed in the Experiments section. Furthermore, we also construct multiple noisy domains by manually adding different magnitudes of Poisson noise.

Label maps of each patient were created to represent the ground truth organ occupancy, each containing nine cardiac substructures which were manually delineated by one physician, as described elsewhere[24]. Briefly, the nine substructures were: Aortic Valve (AV), Left Anterior Descending (LAD), Tricuspid Valve (TV), Mitral Valve (MV), Pulmonic Valve (PV), Right Atrium (RA), Right Ventricle (RV), Left Atrium (LA) and Left Ventricle (LV).

In the preprocessing, images and label maps of each patient were resampled to a 3D spacing of 2mm*2mm*2mm using a linear and nearest-neighbor interpolator, respectively. For image cropping, a 3D bounding box was first determined based on the manual label maps. Then the bounding box was expanded symmetrically in each direction to a size of 100*100*100 voxels, to become the final cropping region. The intensities of the CT images were normalized to a range of 0-1, to be the inputs of segmentation models.

We carefully designed and collected the above datasets from different domains with the aim of gaining an initial understanding of DL segmentation robustness. However, initial experiments showed that the scanner parameters FOV, slice thickness, and reconstruction kernel did not significantly affect segmentation performances (will be shown below). Therefore, these parameters were not considered in the training of the image domain shift detector. Instead, the $RPV$ domain was replaced by $RPV_{Noisy}$ domain, where we randomly add different magnitudes of noise to the images. The entire dataset of 181 images was split by a ratio of 2:1, where 120 patients were used for the training and validation, 61 patients were held out for testing.

## B. Image Domain Shift Detector

In this study, an image domain shift detector was proposed to detect and measure image appearance shift, image intensity shift and image noise level change.

**Image Appearance Shift**. The performance of segmentation models usually degenerates when the input image domain shifts. We attempted to quantify such shift and thus find it's correlation with regard to segmentation accuracy. While systematical changes in image appearance like noise level or resolution that affected by scanning parameters are straightforward to quantify (e.g., by accessing DICOM Metadata), the changes caused by patient factors – such as differences in anatomy, disease, or patient population – are not. [19]In this work, we leveraged a Denoising Autoencoder (referred to as DAE-image) to project CT images to a latent low-dimensional space as feature representations. Compared to generic autoencoders, DAE prevents the network from learning the identity function by randomly adding noise to the network input during the training. The learned image latent representation can be used to evaluate the magnitude of image domain shift.

We trained a 3D DAE (referred to as "DAE-img") from scratch [24]using a training dataset with 121 patients. The encoder comprises 4 convolutional layers and 9 residual blocks[25] and outputs a latent vector with a dimension of 256, while the decoder had a symmetric structure with the encoder. The Mean Square Error (MSE) was used as the final loss function, formulated as:

$$Loss_{DAE-img} = MSE(I, \hat{I}) = \frac{1}{N \sum_{i=1}^{N}(g_i - p_i)^2}$$

Where $I$ and $\hat{I}$ respectively represent the input and the output CT images of the DAE, $p_i$ is the predicted intensity of voxel $i$ , and $g_i$ is the ground truth intensity of voxel $i$. $N$ is the number of voxels per image.

Given an image $I$, we obtained the encoder output $z = Encoder(I)$. Using $z$ as the latent representation of image $I$, we calculated its cosine similarity with $z_{com}^{i}$ , which is the latent representation of image $I_{com}^{i}$ from the $common$ domain. The averaged cosine similarity was used as a measurement of appearance shift for image $I$.

$$X_{Appr\,(I)} = \frac{1}{n \sum_{i=1}^{n} Cosine(z, z_{com}^{i})} = \frac{1}{n \sum_{i=1}^{n} \frac{z \cdot z_{com}^{i}}{\|z\| \|z_{com}^{i}\|}}$$

Where n is the number of images in $common$ domain. The $X_{Appr}$ were used as an input for the developed regression model. A higher value represents visually closer to the $common$ domain, thus indicating higher segmentation performance.

**Image Intensity Shift.** In previous studies[24], we found that certain DL segmentation models trained with non-contrast images performed poorly on contrast images (DSC<0.2), which indicates that these segmentation models might be susceptible to image intensity change. In this study, we explicitly study the relationship between image contrast variations and segmentation performance. For each image, the average voxel value within a whole-heart mask were calculated. Denote this average voxel value by *h*. Then we used $X_{intensity}$ to represent the intensity deviation of a given image compared to the $common$ domain and formulate it as:

$$X_{Intensity} = h - \frac{1}{n \sum_{i=1}^{n} h_{com}^{i}},$$

where $h_{com}^{i}$ denotes the intensity value of an image $I^{i}$ from the common domain and $n$ denotes the number of patients in the common domain. Intuitively, a higher value of $X_{Intensity}$ represents more intensity deviation from the non-contrast image domain(common), thus could potentially lead to lower segmentation performance.

**Image Noise Level.** The magnitude of the image noise may also affect auto-segmentation performance. To estimate the noise level of an image $I$, we first process the image with the aim of reducing the image noise, by using a median filter with a side length of 3, obtaining a denoised image $\hat{I}$, then we calculate the pixel-wise mean square error between the two images before and after the denoising. We use this error metric to quantify the noise magnitude of a given image:

$$X_{noise} = \frac{1}{n \sum_{i}^{n}(I_i - \hat{I}_i)^2}$$

## C. Shape Quality Estimator

Due to the nature of invariability of cardiac substructures, we can infer that "good quality" segmentations form a certain shape distribution and provide strong prior information. (i.e., each cardiac substructure always stays in a relatively fixed location and shape). Based on the work by Liu et.al.[22], we utilized VAE[26] to learn such a shape distribution. The pipeline is shown in Figure 3. The underlying rationale is that VAE can project labels to a low-dimensional feature space. We trained the VAE using only ground truth segmentation labels, which captured the distribution of normal cardiac shape. At the inference stage, bad segmentations will also be projected into the low-dimensional space (i.e. the normal shape distribution) and still be decoded as a "normally shaped segmentation", thus results in high error between VAE input and output. Such error can represent the deviation of given segmentation results to the normal shape distribution.

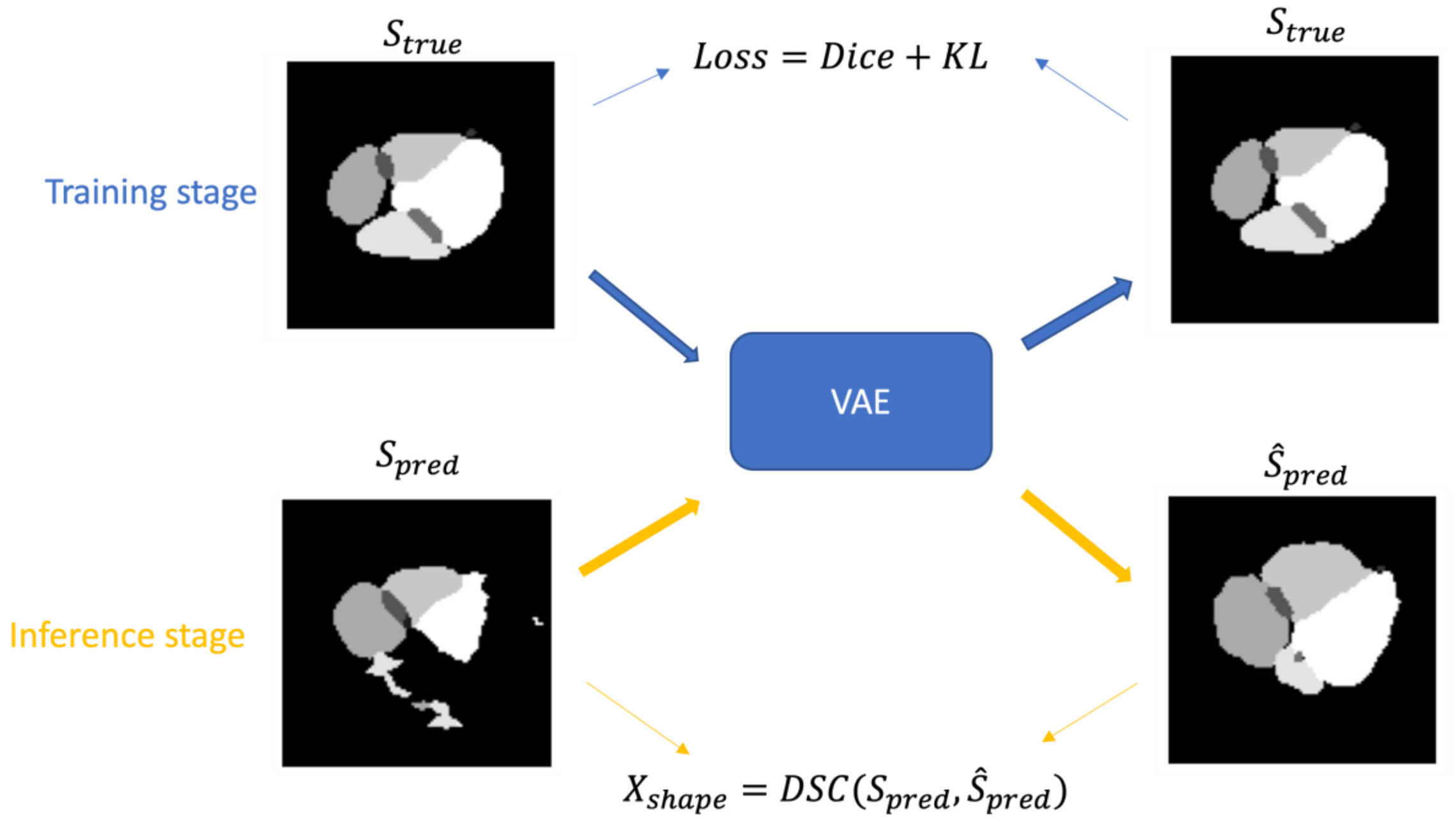

*Figure 3. The pipeline of using a VAE to assess the shape quality of auto-segmentation. The VAE was trained using ground truth labels. At the inference stage, the VAE takes in the results of auto-segmentation and output the reconstructed label maps with "good shape". The DSC overlap between VAE input and output is used as a feature to represent shape quality.*

In this study, we used ground truth labels of cardiac substructure as training/validation data to train a 3D VAE (referred to as "VAE-seg") from scratch based on the implementation of Kingma et.al.[26]. A combination of soft Dice Loss and KL loss were used as the final loss function, formulated as:

$$Loss_{VAE-seg} = Dice(S, \hat{S}) + KL(\mu, \sigma)$$
$$= \frac{1}{L \sum_{l=1}^{L}(1 - \frac{2\sum_i^N p_{li} * g_{li}}{\sum_i^N p_{li}^2 + \sum_i^N g_{li}^2})} - \frac{1}{2}(2\log(\sigma) - \sigma^2 - \mu^2 + 1)$$

Where $S$ and $\hat{S}$ respectively represent the input and the output segmentation map of the VAE-seg, $p_{li}$ is the predicted probability that the voxel $i$ belongs to class $l$, and $g_{li}$ is the ground truth label of class $l$ (binary). $L$ is the number of classes and equals 10 in this study. $\mu$ and $\sigma$ are the output of the encoder, representing the mean and standard deviation of learned Gaussian distribution for each individual input.

At the inference stage, a given auto-segmentation result $S_{pred}$ will be input into the VAE-seg and be reconstructed as the output Ŝ. The difference between $S_{pred}$ and $\hat{S}_{pred}$, measured by Dice Similarity Coefficient (DSC), $X_{shape} = DSC(S_{pred}, \hat{S}_{pred})$ will be used as a feature to represent segmentation shape quality.

## D. Regression model

In the previous sections, for each individual patient case we have extracted domain specific features of CT images and the shape quality of the segmentation output. The target of filling the gap between the image-segmentation pair and the segmentation quality can be formulated as a regression task:

$$Y = G(X_{Appr}, X_{Intensity}, X_{Noise}, X_{Shape})$$

Where $X_{Appr}$ describes the magnitude of image appearance shift, $X_{Intensity}$ describes the magnitude of image intensity shift (cardiac area),$X_{Noise}$ represents the image noise level and $X_{shape}$ describes the quality of the output segmentation shape. $Y$ is the segmentation accuracy, measured by DSC.

We trained a Bagging regressor with an ensemble of 100 linear regressors. For each regressor, the bootstrap sampling rate is 0.9. For the training/validation data of the regression model, the inputs of the $G$ were obtained through the image domain shift detector and shape quality estimator that are described in above sections. The ground truth of output $Y$ were obtained by calculating the DSC between the output of a given segmentation model and its corresponding ground truth labels.

## E. Evaluation of the Quality Assurance framework

**Segmentation models**. In total 19 segmentation models were developed and trained. As the goal of the work is to develop a QA approach for auto-segmentation algorithms, we developed a set of segmentation models to evaluate the QA approach to assess performance with several different segmentation approaches. Developing our own models allowed us to control over the structure, size, and complexity of the models in a known fashion. 60 additional patients were also collected from the same domain as the *common* domain for training and validation of these in-house segmentation models. There have been extensive studies showing that Vision Transformers are more robust to image corruption and domain shift than CNN[27,28], which is largely benefited by its unique self-attention-like architecture[29]. In this work, both CNN-based and transformer-based segmentation models were investigated. For CNN models, we used an in-house algorithm[24] that utilized a combination of ResNet[25] and U-Net[30]. We also investigated the variation of the model by using different loss functions, data augmentation and network sizes, resulted in 18 trained CNN models. Considering the rapidly increased popularity of Transformer[31] and its success in vision tasks, we also trained a 3D model with a Transformer Encoder and a regular CNN decoder. (See the Appendix I for implementation details). The per-case segmentation accuracy was measured by the average DSC of the nine cardiac substructures.

**Regression model evaluation.** It has been shown that different segmentation models may have very different performance on various input image domains, the entire QA pipeline was tested on 19 segmentation models $F^{i} \ ^{(i=1,2..19)}$ for the evaluation of generalization ability. For each segmentation model $F^{i}$, we trained a new regressor $G^{i}$ using the corresponding $X^{i}_{shape}$ and $Y^{i}_{true} = DSC(S^{I}_{pred}, S_{true})$ then test $G^{i}$ on the test dataset. The performance of each $G^{i}$ was evaluated by mean absolute error (MAE) between true DSC and predicted DSC. The final evaluation metric was calculated as the average MAE across all regression models.

$$MAE = \frac{1}{M}\sum_{i=1}^{M}\frac{1}{N}\sum_{j=1}^{N}\left|Y^{ij}_{true} - Y^{ij}_{Pred}\right|,$$

where $M$ denote the number of segmentation models to be evaluated and $N$ denotes the size of the test dataset.

# III. Experiments

## A. The effects of reconstruction protocols on segmentation performance

In this study, we systematically investigated the effects of CT reconstruction protocols on segmentation performance, including noise level, slice thickness, FOV and reconstruction kernel. We conducted controlled experiment by changing one parameter at a time based on the settings of the $common$ domain and applied the new settings on all the 20 patients from $RPV_{base}$ domain, generating a new domain.

**Noise level.** Noise in CT images are mostly Poisson distributed introduced by fluctuation in photon counting. The signal to noise ratio (SNR) of a CT image is usually affected by pitch and mAs. Ideally, the images of the same patient under different scanning parameters are desirable, however, such “paired images” are hard to acquire in the clinic. In this study we manually added Poisson noise directly following the equation:

$$\hat{I} \sim f\left(k; \lambda = \frac{I_i}{n}\right) * n$$

For each pixel location $i$, we draw a sample $\hat{I}_i$ from a Poisson distribution $f$ with a mean of $\frac{I_i}{n}$, where $n$ is the noise level, ranging in $\{0.01,1,2,4,6,8\}$.The image $I$ was transformed to be discrete and ranged in $[0,255]$. Figure 4 shows example CT images after adding different levels of noise to the same image.

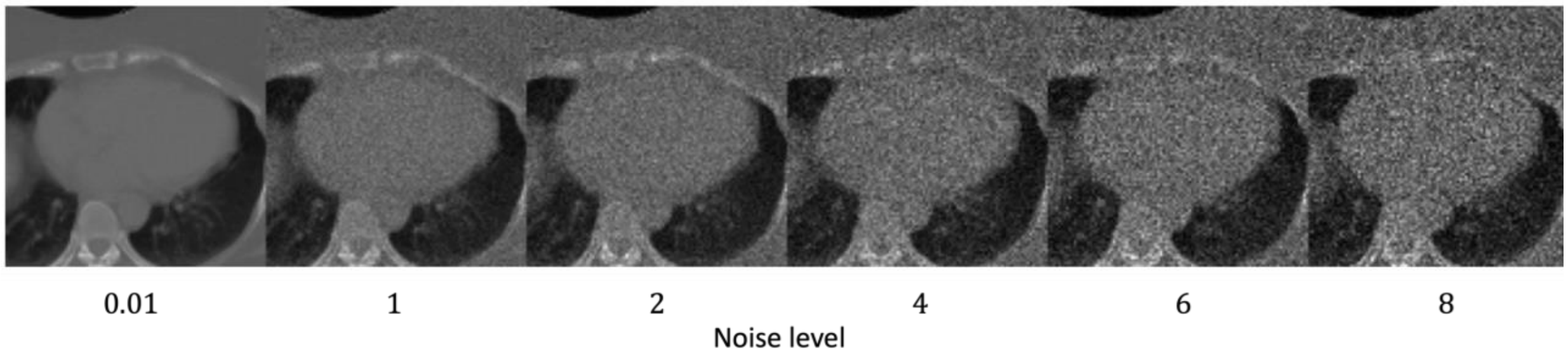

*Figure 4. Example images of one patient from $SV_{base}$ domain, after adding different levels of Poisson noise.*

For the robustness evaluation of segmentation models, at each noise level $n$, we applied noise to the entire $RPV_{base}$ domain, generating a new domain $RPV^{n}_{Noisy}$

**Slice thickness and FOV.** Intuitively, these two parameters both affect image quality due to different spatial resolution. Different slice thicknesses of 1.5mm,2mm,4mm and 5mm were investigated. The smaller thickness yields better image quality. The FOV of 300mm*300mm, 400mm*400mm and 500mm*500mm were also investigated. The dimension of output image slice is fixed to be 512*512, thus smaller FOV yields higher spatial resolution. These images were also preprocessed to a voxel spacing of 2mm*2mm*2mm through a linear interpolator.

**Reconstruction Kernel.** It is common for clinics to use different reconstruction kernel to fit a particular clinical scenario or physician preference. In this work, different built-in reconstruction kernels of a Siemens CT scanner were investigated, including BF-39, BL-57, BR-32, BR-49, BR-62, QR-32 and QR 66. Utilizing the built-in reconstruction software of the scanner, each time we reconstruct new CT images from the raw data of all 20 patients using one selected kernel.

Each reconstructed domain was automatically segmented using the segmentation model and the output was compared to ground truth delineation by DSC as the segmentation accuracy metric. We performed ANOVA test on the DSC results among datasets, to evaluate if there is a significant change caused by a certain scanning parameter.

### B. Evaluation of Compression Models for the Image Domain Shift Detector

In this study, in addition to the DAE we also experimented on the VAE, Gaussian mixture VAE (GM-VAE), Vector Quantized VAE (VQ-VAE) to extract latent image features.

**Variational Autoencoder**. VAE assumes that the data follows a certain underlying parametric probability distribution and attempts to learn the parameters of the distribution. Here we assume the data follows a standard multivariate Gaussian distribution, and the encoder outputs the mean and variation vector of it. The implementation of the encoder and decoder were the same as the DAE.

**Gaussian Mixture VAE.** Given the fact that the image data comes from multiple domains, it makes more sense to assume the data follows a more complicated distribution, e.g., Gaussian Mixture Model. The GMVAE can cluster the similar-looking images in an unsupervised approach[32]. We implemented the GM-VAE based on Dilokthanakul et.al[32]. The number of Gaussian components was 8.

**Vector Quantized VAE.** Compared to VAE, the encoder of VQ-VAE outputs discrete vectors and the prior is learned rather than pre-determined. We implement the VQ-VAE based on Van Dan Oord et.al.[33] with a latent dimension of 216 and a trainable dictionary containing 2048 vectors.

### C. Comparison of the regression methods.

In this study we compared different regression techniques, including naïve linear regression, Bagging, Gaussian Process Regression (GPR), Random Forest (RF) and Gradient Boost Tree (GBT)

**Bootstrap Aggregating (Bagging).** The idea of bagging is to subsample the dataset to train a weak regressor each time, and then aggregate them to make predictions. Here we used linear regression as the basic regressor. The number of regressors was 100 and the subsample rate was 0.9

**Gaussian Process Regression.** GPR is nonparametric and calculates the probability distribution over all admissible functions that fit the data. Given a specified prior, GPR calculates the posterior using the training data, and computes the predictive posterior distributions on test data. One advantage of GPR is that besides predicting exact values, it also provides uncertainty on predictions (usually in the form of 95% confidence interval). In this study, we trained a Gaussian process regressor using a kernel of the combination of RBF with an auto-tuned length scale, and a white kernel with auto-tuned noise level.

**Random Forest.** RF is an ensembled estimator that fits multiple classifying decision trees on various sub-samples of the dataset and uses averaging to improve prediction accuracy and alleviate overfitting. In this study, we trained a random forest regressor with an ensemble of 100 basic decision trees and other default settings using the package Scikit-learn 1.1.3.

**Gradient Boost Regression**. GBT learns weak regressors and aggregates them in a sequential and weighted manner. In this work, we trained a gradient-boosting regressor using 100 weak estimators and a maximum depth of 3.

### D. Comparison of overall QA methods

In this work, our proposed QA framework was compared against two other methods for QA of auto-segmentation systems, CNN based regression and using only the shape quality estimator as proposed by Liu et.al [22]

For Liu's method, we implemented the algorithm by using $X_{Shape}$ as the sole input of the regression model. Based on the work of Chen et.al[34] and Robinson et.al.[35], we develop a CNN which takes in an image and segmentation pair and output the predicted DSC. We discretize the output DSC with a step of 0.01, thus transforming the regression problem into a pseudo classification problem, where the output has 100 classes (i.e., 0.00, 0.01, 0.02, … 0.99). The input image and segmentation were separately input through 6 convolutional layers and then got concatenated. After concatenation, nine convolutional layers followed and output a 100-D vector, based on which the predicted DSC was obtained via $argmax$ operation. The same training and testing dataset were used for the CNN training and testing.

## IV. Results

In this section, we will demonstrate the segmentation performance of the 19 models tested on the benchmark dataset. Then we will individually demonstrate the effectiveness of each input features $X_{Appr}, X_{Intensity}, X_{Noise}, X_{Shape}$ of the regression model by showing the correlation between them and the true DSC. Intuitively, a higher correlation indicates that the feature captures more important information that could affect the segmentation performance, thus making the regression task easier. In the end, the performance of the final regression model is shown, along with the comparison results of different regression methods.

## A. Segmentation Performance on the Benchmark Dataset

**Segmentation Performance Across Image Domains.** Each segmentation model was tested on the benchmark dataset including one $common$ domain and five $uncommon$ domains. Taking two segmentation models as examples, their performances in terms of DSC are shown in Figure 5. The selected CNN-based model showed a significant DSC drop on the $CE$ (Wilcoxon, $p < 1.e^{-9}$) and $RPV_{noisy}$ (Wilcoxon, $p < 1.e^{-3}$) domain, compared to the $common$ domain. The transformer showed a DSC decrease of 0.04 (Wilcoxon, $p < 1.e^{-2}$) on the $AV$ domain and was relatively more robust on other image domains.

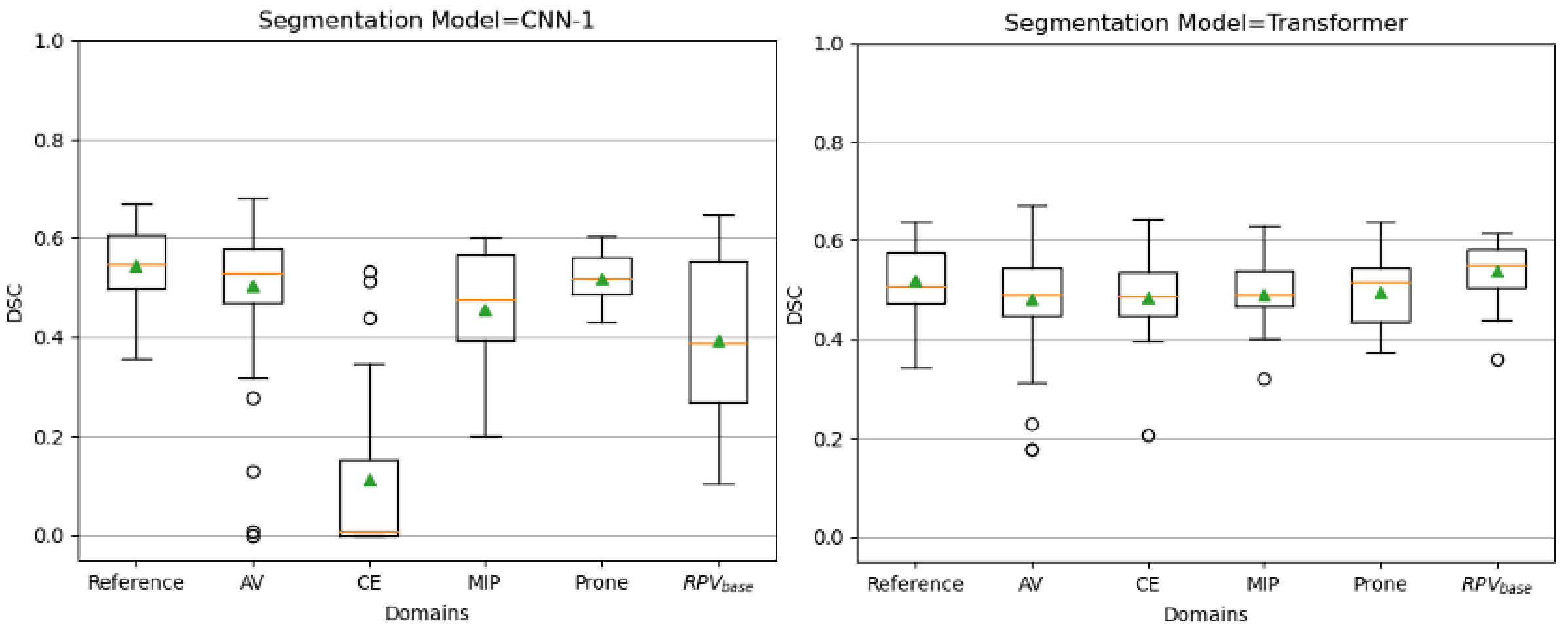

(a) (b)

*Figure 5. The segmentation performance of (a) one CNN-based model and (b) one transformer-based model when tested on different image domains, measured by DSC*

The detailed test results on $CE$ (contrast-enhanced) domain are shown in Figure 6, which shows the relative performance change for different segmentation models on this domain, measured by the difference of the mean DSC on the $Common$ and $CE$ domains. A positive DSC change implies worse accuracy for the $CE$ domain with this model. The CNN-based models all showed significant performance drop ranging from 0.19 to 0.47 (Wilcoxon, $p < 6.e^{-5}$). While the mean DSC drop for the Transformer is 0.03 (Wilcoxon, $p = 0.20$).

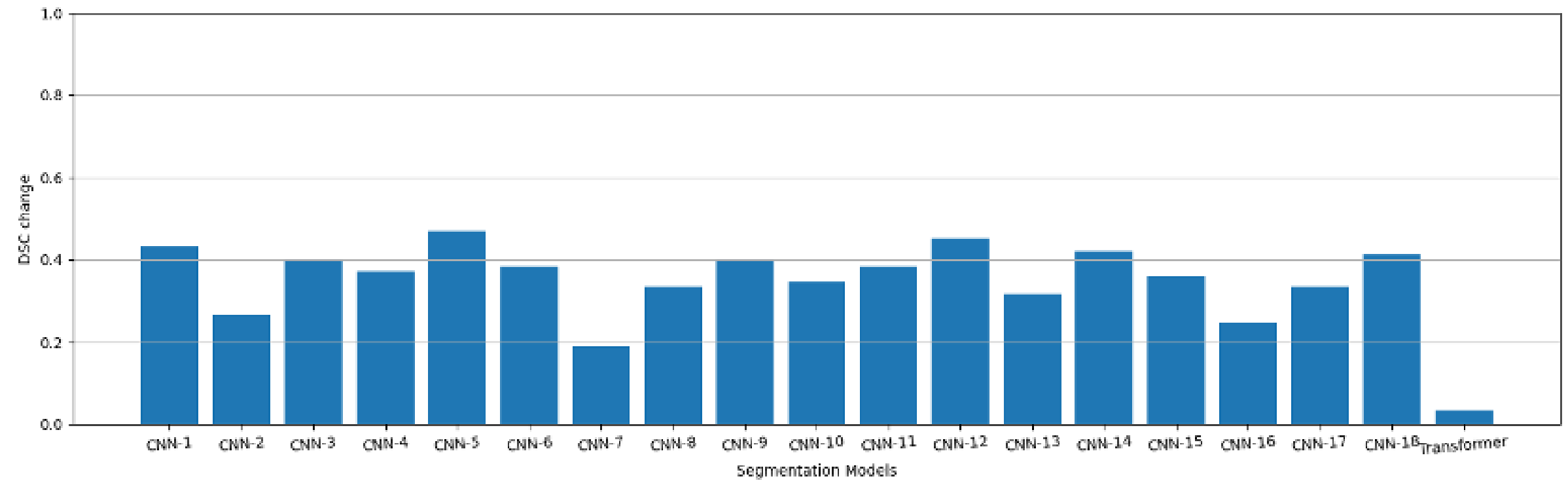

*Figure 6. The performance change of different segmentation models when tested on contrast images, measured by the difference of the mean DSC of the images from $Reference$ domain and $CE$ domain.*

**Segmentation Performance using Different Scanning Parameters**. Experiments have shown that the $slice\ thickness$, $FOV$ and $reconstruction\ kernel$ did not have a significant effect on the performance of all the 19 segmentation models (ANOVA, $p > 0.42$). Detailed test results can be found in the Appendix. To alleviate overfitting by constraining the number of variables, we did not include these scanner parameters as factors when building the regression model.

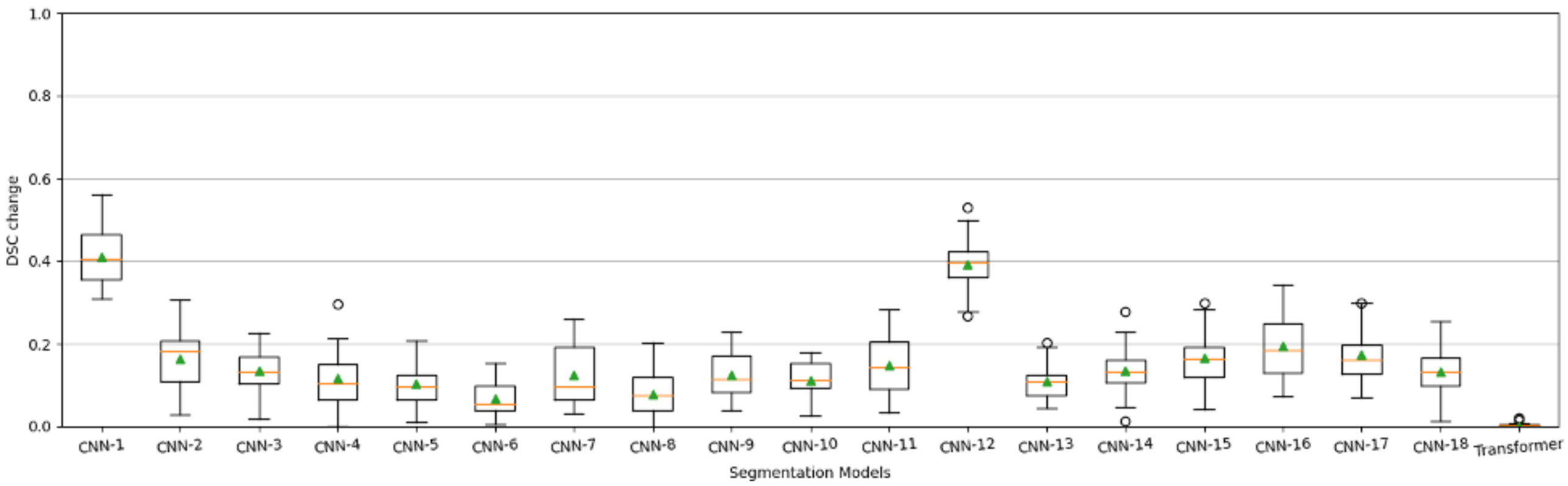

*Figure 7. The performance change (measured by DSC) of 19 segmentation models. For each model, the performance change is measured by the decrease of mean DSC between the test results on $RPV_{base}$ and $RPV_{Noise}^{max}$.*

The performances of different segmentation models on the images with different levels of image noise are shown in Figure 7. For each model, the performance drop is measured by the decrease of mean DSC between the test results on $RPV_{Base}$ domain and $RPV_{Noise}^{n=max}$, where $n$ is the maximum magnitude of noise(i.e., $n = 8$). The image noise level was found to have a significant effect on all CNN based segmentation models (ANOVA, $p < 0.01$). When 20 test images were added with a maximum level of noise, CNN-based models demonstrated DSC drops, ranging from 0.07 (Wilcoxon, $p = 2.e^{-5}$) to 0.41(Wilcoxon, $p = 2.e^{-6}$)(Figure 8 (a)). However, the transformer did not show significant performance change when dealing with noisy images (ANOVA, $p = 0.99$), as shown in Figure 8 (b)

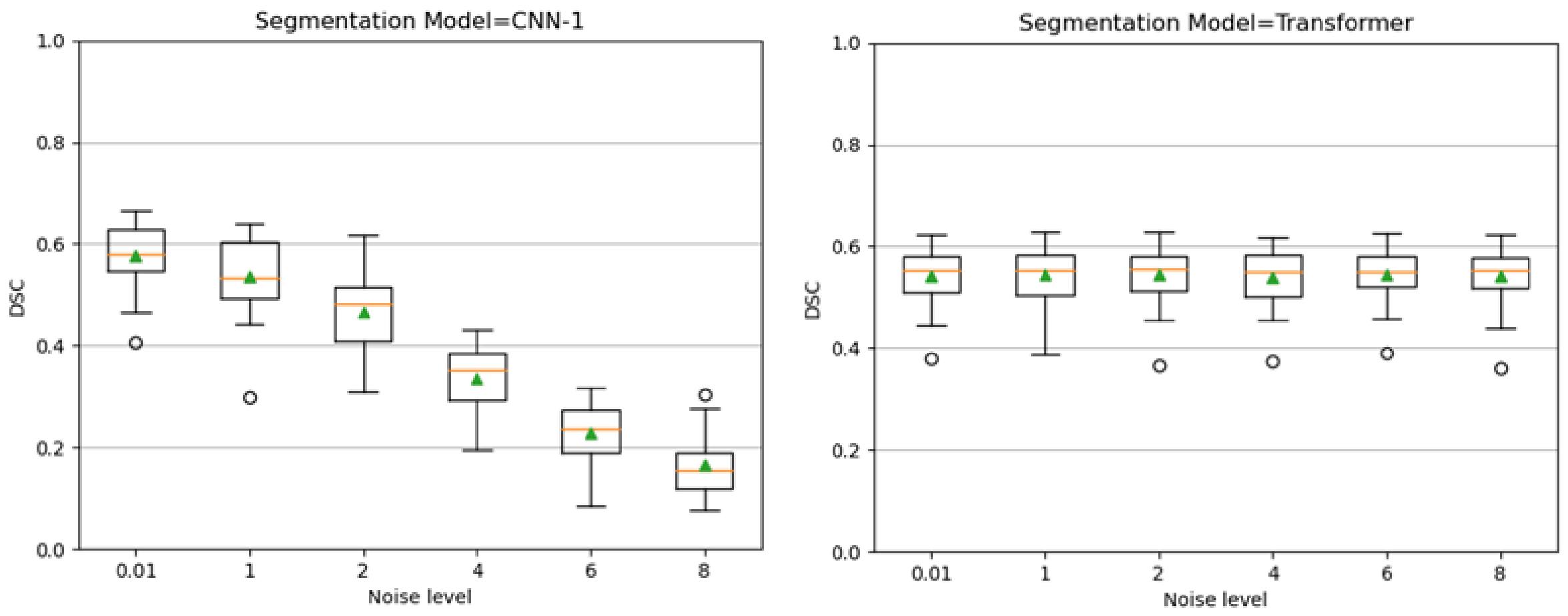

(a) (b)

*Figure 8. The segmentation performance of (a) one CNN-based model and (b) one transformer-based model when tested on images with different levels of noise, measured by DSC. The yellow line in boxplot represents medium and green triangle represents mean value.*

## B. Image Domain Shift Detector

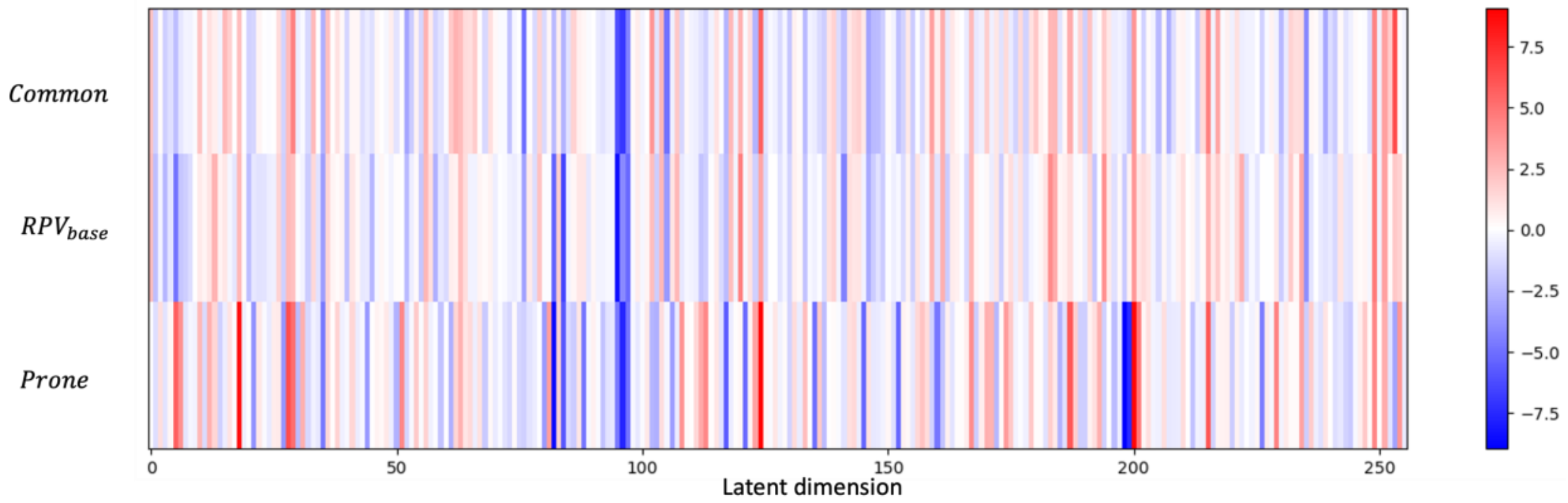

*Figure 9. Mean latent vectors for the DAE-image, averaged over all patients, from three domains: $Reference$, $RPV_{base}$ and $Prone$. For each domain, the averaged vector across all patients is shown.*

**Image Appearance Shift.** Figure 9 shows the latent features for the image domain shift detector by DAE-image from three domains:$Common$, $RPV_{base}$and $Prone$. For each domain, the feature averaged across all patients is shown. We can see that the latent vector of the $Common$ domain and $RPV_{base}$ are visually more similar than the $Prone$ domain, which is consistent with the fact that $Common$ and $RPV_{base}$ are from the same domain. The results indicate that the latent feature vector learned by DAE has the capability to distinguish various image domains.

Figure 10 (a) shows the relationship between $X_{Appr}$ and $Y_{true}$ (DSC). Here we show the data points from the entire dataset regardless of training or testing. Note that the $Y_{true}$ is dependent on the segmentation model. When using a CNN-based segmentation model, the Pearson correlation was 0.278 ($p < 3.e^{-4}$). For a transformer-based segmentation model, the Pearson correlation was 0.415 ($p < 2.e^{-8}$). Notice that in the CNN results, the majority of the samples with low DSC (<0.1) are contrast patients (shown as crosses in Figure 10(a), upper panel). In such cases, the feature $X_{Intensity}$ was dominating, thus resulting in a lower Pearson correlation score. However, the transformer-based segmentation model is less sensitive to image intensity (shown below), thus

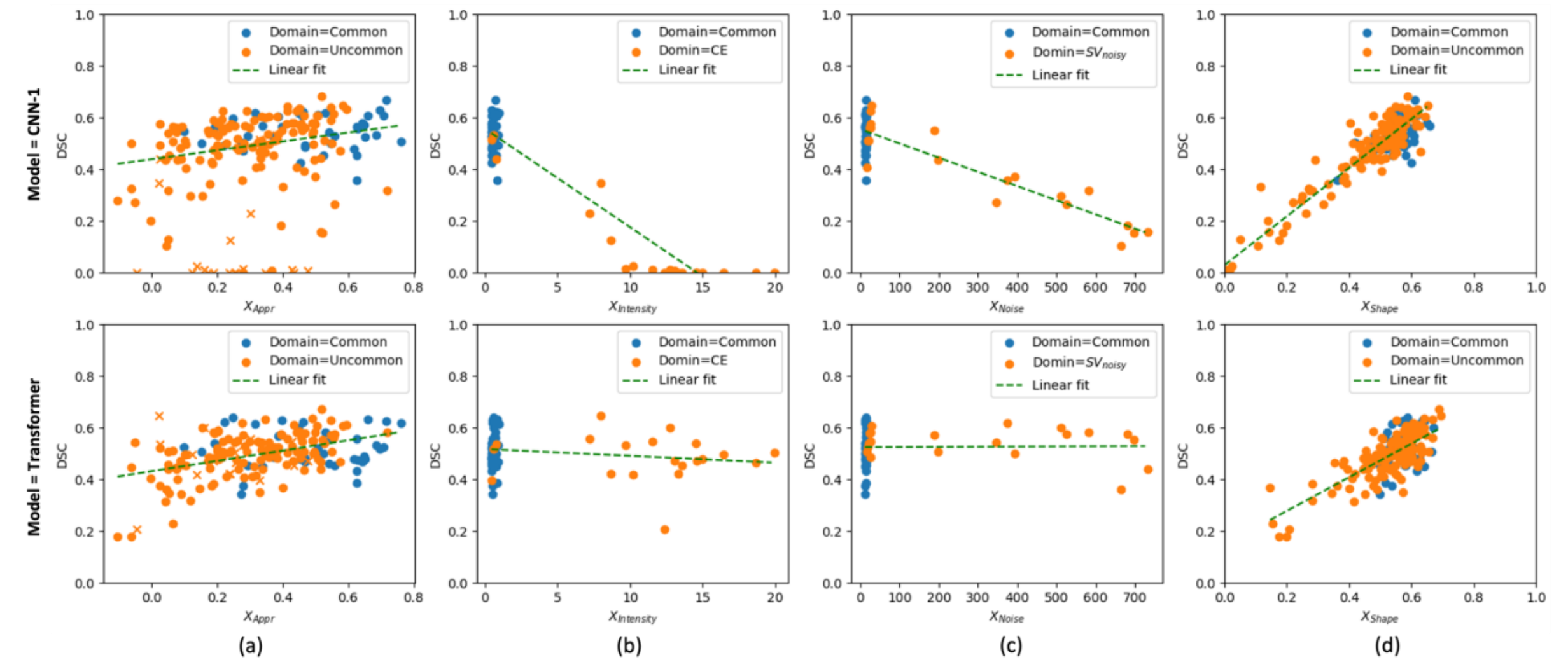

*Figure 10. The relationship between four features $X_{Appr}, X_{Intensity}, X_{Noise}, X_{Shape}$ and $Y_{true}$ (DSC), shown in subfigure (a), (b), (c), (d) respectively. Results of two example segmentation models: one CNN-based model and one transformer-based model are shown in the first and second row respectively. The blue dots represent the common domain, and the orange dots(cross) represents other uncommon domains (AV, CE, MA, Prone, $SV_{noisy}$). The green dashed line is a linearly regressor fitted by the shown datapoints. In subfigure (a), datapoints from the $CE$ datasets are shown as crosses.*

showing a stronger linear relationship between $X_{Appr}$ and $Y_{true}$.

**Image Intensity Shift.** As shown in Figure 10(a), it is apparent that the feature learned by DAE-image, $X_{Appr}$, is not sensitive to image intensity shift. To compensate for this, we explicitly used $X_{Intensity}$ to represent the heart area intensity shift of a given image compared to the common domain. Figure 10(b) shows the relationship between the feature $X_{Intensity}$ and true DSC. In order to isolate other sources of domain variation, we only include the $common$(non-contrast) and $CE$ domain in this experiment. When using a CNN-based segmentation model, the Pearson correlation was -0.946( $p < 2.e^{-29}$). For a transformer-based segmentation model, the Pearson correlation was -0.191 ($p = 0.15$). The results demonstrates that if a segmentation model is susceptible to image intensity change, the feature $X_{Intensity}$ captures a strong correlation between intensity shift and performance change.

**Image Noise Level.** Figure 10(c) shows the relationship between the feature $X_{Noise}$ and true DSC. To limit the number of domain variation sources, we only include the $common$ and $RPV_{noisy}$ domains. When using a CNN-based segmentation model, the Pearson correlation was -0.874 ($p < 2.e^{-19}$). For a transformer-based segmentor, the Pearson correlation was 0.015 ($p = 0.91$). As shown in section A, the transformer is relatively more robust to noise compared to CNN, which explains the lower Pearson correlation score for the transformer. However, given a segmentation model that is susceptible to image noise, the feature $X_{Noise}$ is able to quantify the noise level and establish a linear relationship between it and the segmentation performance.

## C. Shape Quality Estimator

The feature $X_{shape}$ attempts to evaluate the shape quality of the auto-segmentation results. Figure 10(d) shows the relationship between a feature $X_{shape}$ and true DSC. When using a CNN-based segmentation model, the Pearson correlation was 0.953 ($p < 2.e^{-88}$). For a transformer-based segmentation model, the Pearson correlation was 0.708 ($p < 5.e^{-27}$). From the above experiments, the feature $X_{shape}$ demonstrated a strong linear relationship with the segmentation performance for both model architectures.

## D. Performance of the Regression Model

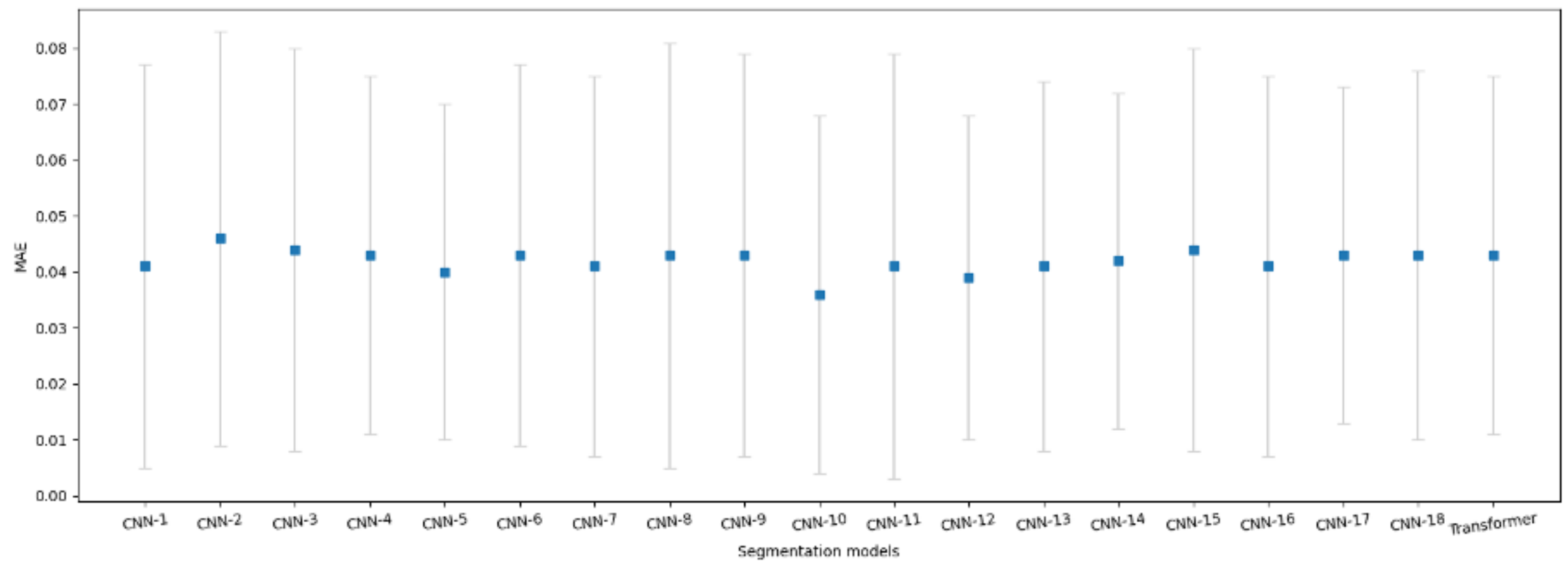

*Figure 11. The test results of the regression models. Each regression model was trained using the data from a certain segmentation model. The regression performance was measured MAE with standard error.*

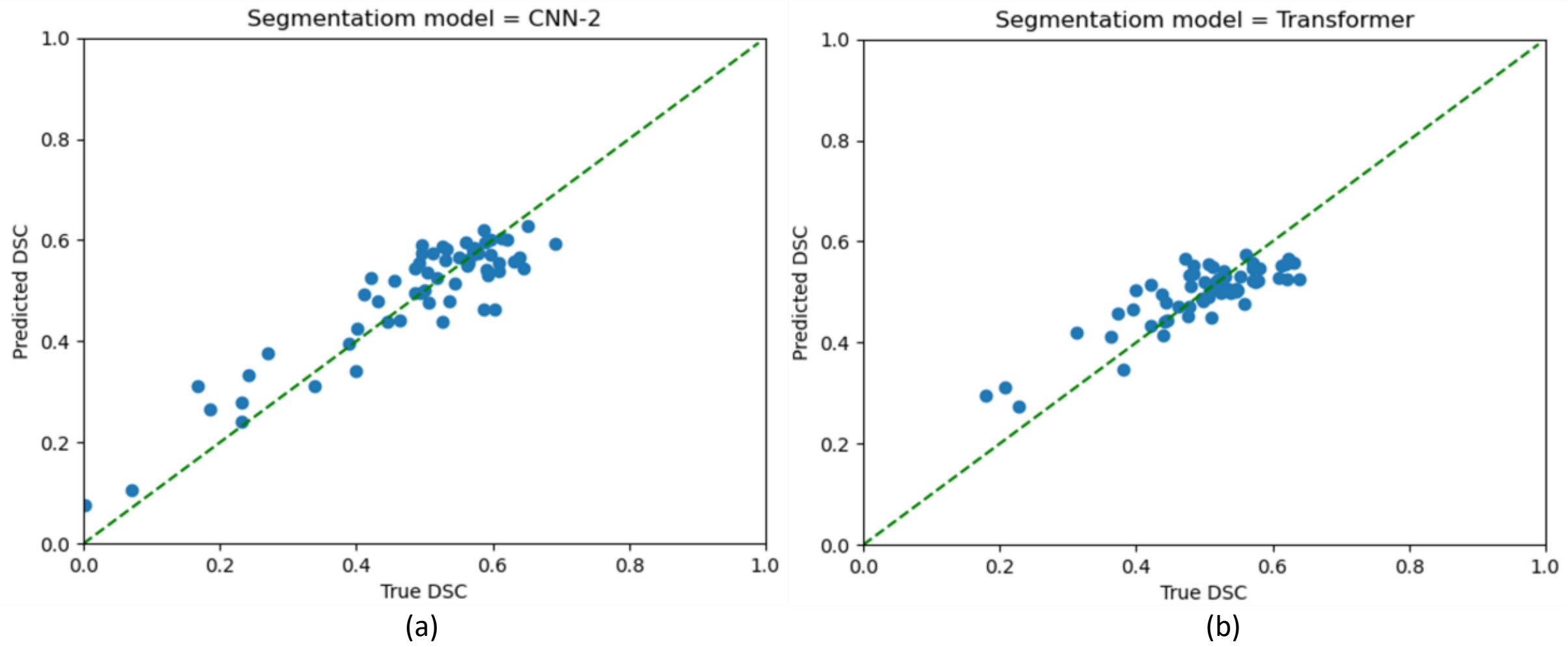

*Figure 12. The regression models prediction on individual test samples. Two example regression models were trained using the data from (a) one CNN-based segmentation model, (b) one transformer-based segmentation model. The green dashed line is the ideal performance with 100% accuracy.*

When tested on 60 patients, the trained Bagging Regressors predicted the segmentation accuracy (DSC) with the mean absolute error (MAE) in a range of 0.036 to 0.046 across 19 segmentation models. The average MAE is $0.042 \pm 0.002$. When selecting the DSC of 0.4 as an empirical threshold of 'good' vs. 'poor' segmentation performance for a segmentation system, the proposed method achieves a classification accuracy of $0.92$ to $1.00$ among 19 segmentation models, with an average accuracy of 0.97. The MAE performances of the regressors are shown in Figure 11. Choosing one CNN segmentation model and one transformer segmentation model as examples, Figure 12 shows the regressor's predictions on individual test samples.

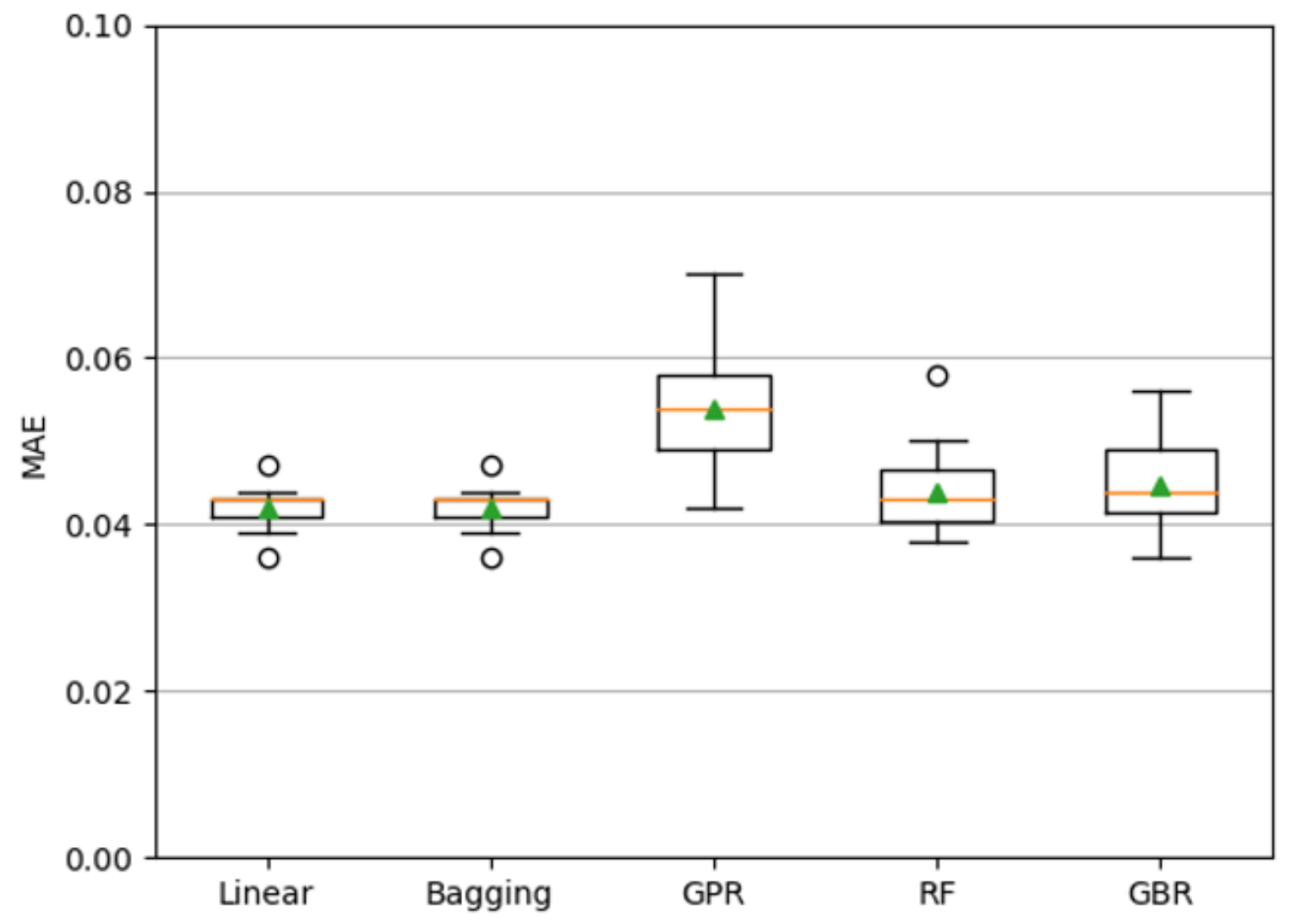

*Figure 13. Results of the comparison between different regression methods, measured by average MAE. The yellow line in boxplot represents medium and green triangle represents mean value.*

Figure 13 shows the regression results (measured by MAE) of different regression methods. The lower MAE indicates better performance. The performance of naive linear regression and bagging were similar (Wilcoxon, $p > 0.22$), both with a MAE of $0.042 \pm 0.002$, which was an average value across different segmentation models. Gaussian Process Regression (GPR) showed a higher MAE of $0.054 \pm 0.007$. Random Forest (RF) and Gradient Boost regressor (GBR) separately achieved the average MAE of $0.044 \pm 0.005$, and $0.045 \pm 0.003$.

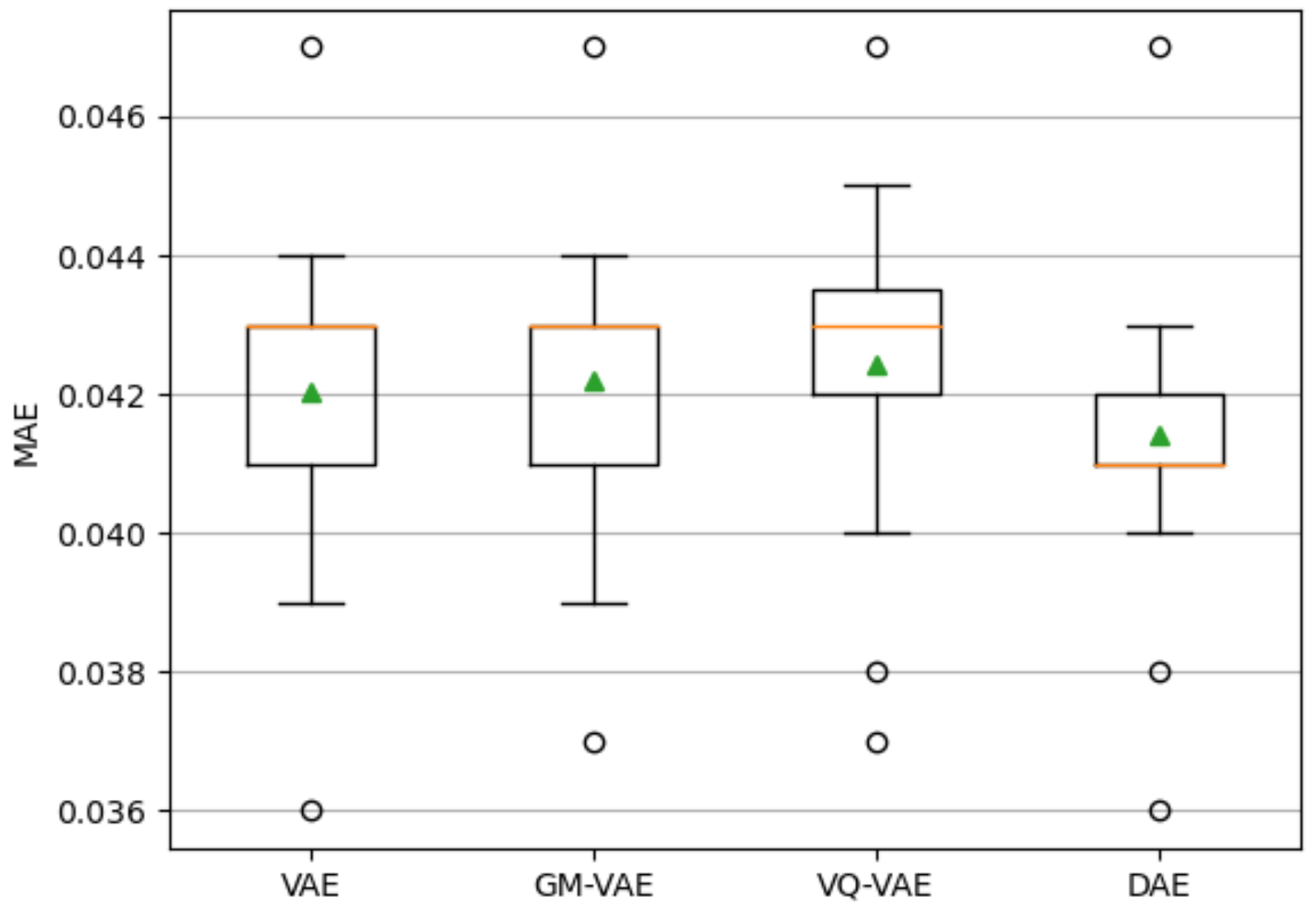

*Figure 14. Comparison of methods to quantifying image appearance shift: Variational autoencoder, Gaussian Mixture VAE, Vector quantized VAE and Denoising autoencoder. The Mean absolute error (MAE) of the final regression model is shown. The yellow line in boxplot represents medium and green triangle represents mean value.*

**E. Results of comparative experiments**

**Comparison of Methods in image representation learning.** We investigated different approaches to quantify image appearance shift and compared their performance in terms of final regression error. When tested on 19 segmentation models, the MAE results of each method were shown in Figure 14. The results when using VAE and GM-VAE were statistically similar (Wilcoxon, $p = 0.08$), while DAE had a significantly lower MAE than three other methods (Wilcoxon, $p < 3.e^{-3}$).

**Comparison of QA methods.** Here we have demonstrated the comparison results among three QA methods, our proposed framework, Liu's method and CNN-based regression. Tested on 19 segmentation models, the MAE results of each method were shown in Figure 15.

Compared to our proposed method, Liu et.al.'s method[22] had a higher MAE of $0.045 \pm 0.003$ (Wilcoxon, $p < 4.e^{-6}$). The CNN based regression achieve a MAE of $0.096 \pm 0.017$(Wilcoxon, $p < 4.e^{-6}$), which is the highest among the three methods.

# V. Discussion

Among other related works, this work is the first to develop a real-time QA system for CT image cardiac substructure segmentation. Overall, the proposed method is able to accurately predict segmentation quality (as measured here by DSC) across multiple evaluated segmentation algorithms, with MAEs in a range of 0.036 to 0.046. Instead of using the public dataset, which usually contains images carefully curated from study protocols and patient cohorts, in this work we carefully designed a benchmark dataset that comprises images from exhaustive domains relevant to our clinic to best simulate the clinical setting and test the segmentation robustness from extensive aspects. The benchmark not only serves the purpose of helping to train the QA system, but it can also be used in commissioning or system selection, e.g., to rank several segmentation algorithms or options against each other before they are deployed. The test results of multiple segmentation models on the benchmark dataset showed that the tested models with both network architectures were susceptible to patients with anatomical variation. For contrast and noisy images, the CNN-based models showed significant performance degradation (contrast: $p < 6.e^{-5}$ , noisy:$p < 2.e^{-5}$ ) while the transformer-based model was not affected (contrast:$p > 0.2$, noisy: $p > 0.99$).

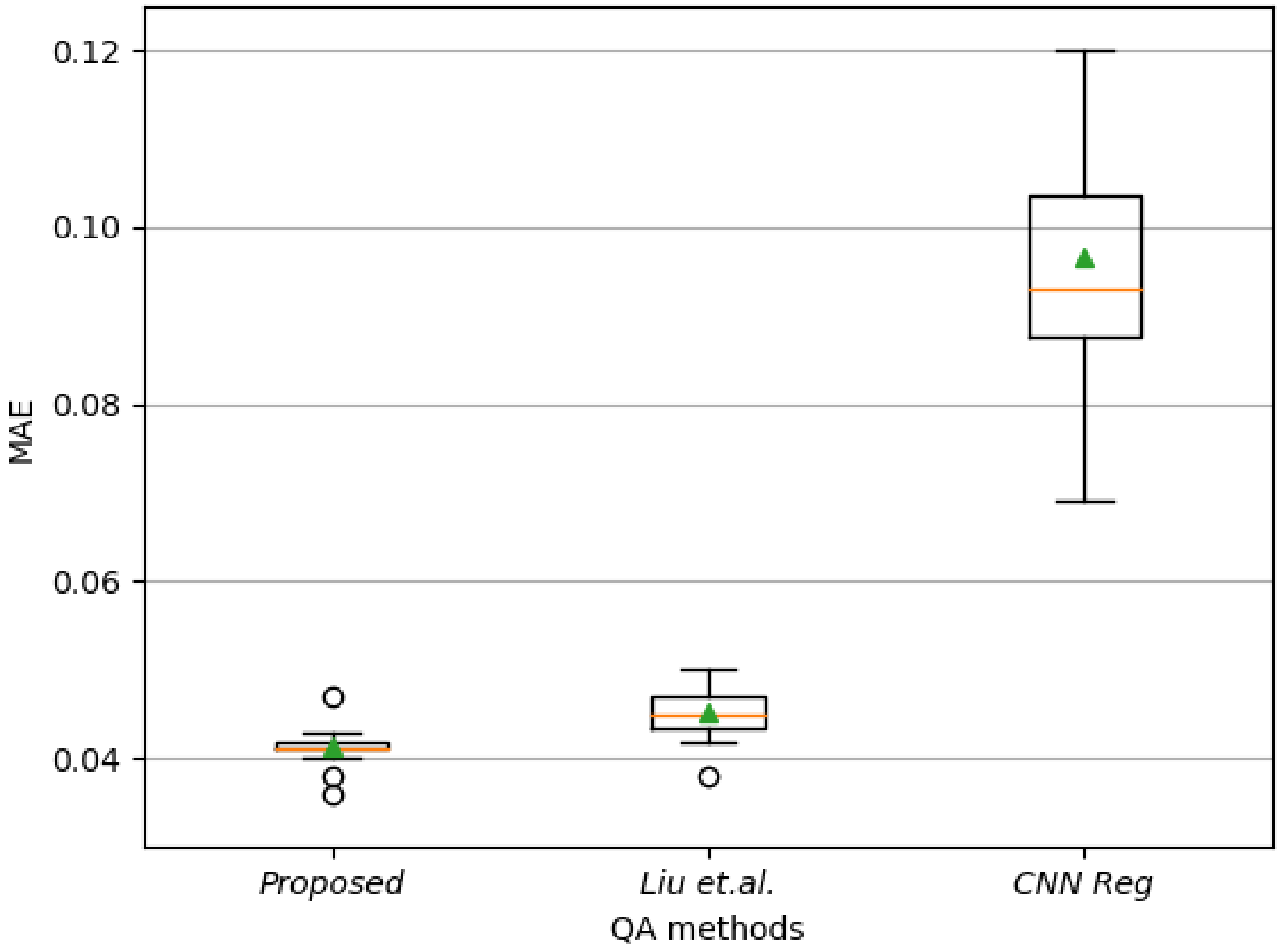

*Figure 15. Results of the comparison between different QA methods, measured by average MAE, tested on 19 segmentation models. The three methods are (1) Our proposed framework. (2) Liu's method which only used the shape quality estimator. (3) Direct regression from image-segmentation pairs using CNN. The yellow line in boxplot represents medium and green triangle represents mean value.*

The image latent features $X_{Appr}$ that were learned via an unsupervised manner have the ability to distinguish between image domains. The similarities between the latent representations of $common$ domain and simulated clinic domains were found to be correlated to segmentation performance. Two hand-engineered image features: $X_{Intensity}$ and $X_{Noise}$ were found to be strong representatives of segmentation performance degradation on contrast or noisy CT images.$X_{Shape}$ In this study, we leveraged the fact that the shape and location of cardiac structures are relatively deterministic (compared to – for example – tumors) amongst a population and used a VAE to learn such a shape representation. Experiments showed that the trained VAE was able to reconstruct a realistic looking segmentation from a failed auto-segmentation. The feature $X_{shape}$ , which calculates the similarity (i.e. DSC) between the VAE's input and output, was found strongly correlated to actual segmentation accuracy.

The output of the regression model provides per-case prediction, which can be used with a threshold (e.g. DSC<0.4) to detect out-of-distribution performance. A threshold should be selected and tuned to the clinical problem, but here we showed good classification performance with an arbitrarily-selected threshold. The proposed QA system can also be utilized to examine input images and output segmentation in batches, and thus could be used to monitor for any long-term data drift. For example, if the predicted DSCs over the recent 50 patients have significantly dropped compared to the DSC of the common dataset, this may suggest a review for potential systematic changes in scanner protocols or patient cohorts and re-evaluate the segmentation model.

Among other related works, Soin et al.[20] utilized VAE and imaging metadata to propose a drift score that is correlated to ground truth model accuracy in an x-ray image classification task. However, they only evaluate the proposed pipeline on one classification model, thus lacking the demonstration of generalizability. Their method also requires access to the DL model's raw output, while ours does not. Wang et al.[17] proposed a generative model based QA framework for the task of cardiac MR segmentation and reported a MAE of 0.03 and 0.07 on two public datasets. However, their iterative search could result in slower inference speed. Robinson et al.[35] developed a CNN-based method for real-time regression of DSC from image-segmentation pairs and reported a MAE of 0.03. Liu et al[22]. proposed to estimate segmentation accuracy by learning the shape statistics of ground truth label maps and achieved a MAE of 0.03, 0.08 and 0.05 on three public datasets. Compared to other related works, our proposed QA framework not only achieved similar or better performance but was also tested across multiple segmentation algorithms for the evaluation of generalizability. Apart from that, our work highlighted the importance of a task-specific benchmark dataset. Such a dataset can be used to supplement the typically larger, more general training datasets employed by those developing clinical-grade autosegmentation systems by adding in task- or patient population-specific images and anatomies that are less common. Under such circumstance, we explored and proposed the methodology of customizing a light-weight benchmark dataset, which is not only clinically feasible, but also provides better customization for the image domains that we are interested in. Combining domain knowledge, the proposed framework has potential to be extended to arbitrary clinic segmentation tasks and other image modalities.

One limitation of this study is that we only used Dice coefficient as the measurement of segmentation accuracy. DSC is a good representative of segmentation accuracy on large structures. However, for small structures like cardiac arteries, a small location deviation could result in a low DSC whilst being clinically insignificant[36] and vice versa. Metrics that focus on the clinically relevant tasks of interest may help address this issue[37,38,39], and will be incorporated in the QA system in the future. Another limitation is that the shape assessment of output segmentation via VAE was developed under the assumption that the shapes of the segmentation targets are relatively deterministic. Therefore, this feature might succeed on monitoring the auto-segmentation of organ-at-risk (OAR) but might fail on the segmentation target of a tumor or other pathology.

Future works will mainly involve testing the proposed framework on other (commercial) segmentation algorithms and transferring the framework to other clinical applications.

# VI. Conclusion

In this work, we proposed a quality assurance framework for monitoring the performance of DL segmentation models in radiotherapy and validated it on an example task of cardiac substructure segmentation. The QA framework includes three components: 1) a comprehensive benchmark dataset. 2) Autoencoder based techniques and hand-engineered features for image domain shift quantification and segmentation shape assessment. 3) a regression model that takes in the extracted features and makes clinically acceptable predictions of segmentation performances. The robustness of multiple CNN-based and transformer-based segmentation models was systematically investigated using the QA framework. The CNN-based models showed decreased performance on contrast CT and noisy images while the transformer model showed relatively more robust performance across different image domains.

## Reference

1. Mårtensson G, Ferreira D, Granberg T, et al. The reliability of a deep learning model in clinical out-of-distribution MRI data: A multicohort study. *Med Image Anal*. 2020;66:101714. doi:10.1016/j.media.2020.101714

2. Feng X, Bernard ME, Hunter T, Chen Q. Improving accuracy and robustness of deep convolutional neural network based thoracic OAR segmentation. *Phys Med Biol*. 2020;65(7):7-8. doi:10.1088/1361-6560/ab7877

3. Full PM, Isensee F, Jäger PF, Maier-Hein K. Studying Robustness of Semantic Segmentation under Domain Shift in cardiac MRI. *Lecture Notes in Computer Science (including subseries Lecture Notes in Artificial Intelligence and Lecture Notes in Bioinformatics)*. 2020;12592 LNCS:238-249. Accessed April 5, 2021. http://arxiv.org/abs/2011.07592

4. Glocker B, Robinson R, Castro DC, Dou Q, Konukoglu E. Machine Learning with Multi-Site Imaging Data: An Empirical Study on the Impact of Scanner Effects. *ArXiv*. Published online October 10, 2019. Accessed February 18, 2021. http://arxiv.org/abs/1910.04597

5. Karani N, Chaitanya K, Baumgartner C, Konukoglu E. A lifelong learning approach to brain MR segmentation across scanners and protocols. In: *Lecture Notes in Computer Science (Including Subseries Lecture Notes in Artificial Intelligence and Lecture Notes in Bioinformatics)*. Vol 11070 LNCS. Springer Verlag; 2018:476-484. doi:10.1007/978-3-030-00928-1_54

6. Karani N, Erdil E, Chaitanya K, Konukoglu E. Test-Time Adaptable Neural Networks for Robust Medical Image Segmentation. *Med Image Anal*. 2020;68. doi:10.1016/j.media.2020.101907

7. Zech JR, Badgeley MA, Liu M, Costa AB, Titano JJ, Oermann EK. Confounding variables can degrade generalization performance of radiological deep learning models. *ArXiv*. 2018;15(11):e1002683. doi:10.1371/journal.pmed.1002683

8. Huang Y, Würfl T, Breininger K, Liu L, Lauritsch G, Maier A. Some investigations on robustness of deep learning in limited angle tomography. In: *Lecture Notes in Computer Science (Including Subseries Lecture Notes in Artificial Intelligence and Lecture Notes in Bioinformatics)*. Vol 11070 LNCS. Springer Verlag; 2018:145-153. doi:10.1007/978-3-030-00928-1_17

9. Maier J, Eulig E, Vöth T, et al. Real-time scatter estimation for medical CT using the deep scatter estimation: Method and robustness analysis with respect to different anatomies, dose levels, tube voltages, and data truncation. *Med Phys*. 2019;46(1):238-249. doi:10.1002/mp.13274

10. Hooper SM, Dunnmon JA, Lungren MP, et al. Assessing Robustness to Noise: Low-Cost Head CT Triage. *ArXiv*. Published online March 17, 2020. Accessed May 7, 2021. http://arxiv.org/abs/2003.07977

11. Vandewinckele L, Claessens M, Dinkla A, et al. Overview of artificial intelligence-based applications in radiotherapy: Recommendations for implementation and quality assurance. *Radiotherapy and Oncology*. 2020;153:55-66. doi:10.1016/J.RADONC.2020.09.008

12. Nealon KA, Balter PA, Douglas RJ, et al. Using Failure Mode and Effects Analysis to Evaluate Risk in the Clinical Adoption of Automated Contouring and Treatment Planning Tools. *Pract Radiat Oncol*. 2022;12(4):e344-e353. doi:10.1016/J.PRRO.2022.01.003

13. Kisling K, Johnson JL, Simonds H, et al. A risk assessment of automated treatment planning and recommendations for clinical deployment. *Med Phys*. 2019;46(6):2567-2574. doi:10.1002/MP.13552

14. Claessens M, Oria CS, Brouwer C, et al. Quality Assurance for AI-Based Applications in Radiation Therapy. *Semin Radiat Oncol*. 2022;32(4):421-431. doi:10.1016/J.SEMRADONC.2022.06.011

15. Li K, Yu L, Heng PA. Towards reliable cardiac image segmentation: Assessing image-level and pixel-level segmentation quality via self-reflective references. *Med Image Anal*. 2022;78:102426. doi:10.1016/J.MEDIA.2022.102426

16. Robinson R, Oktay O, Bai W, et al. Real-time Prediction of Segmentation Quality.

17. Wang S, Tarroni G, Qin C, et al. Deep Generative Model-based Quality Control for Cardiac MRI Segmentation.

18. DeVries T, Taylor GW. Leveraging Uncertainty Estimates for Predicting Segmentation Quality. Published online July 2, 2018. doi:10.48550/arxiv.1807.00502

19. Jha AK, Bradshaw TJ, Buvat I, et al. Nuclear Medicine and Artificial Intelligence: Best Practices for Evaluation (the RELAINCE guidelines). *Journal of Nuclear Medicine*. 2022;63(9):1288-1299. doi:10.2967/JNUMED.121.263239

20. Soin A, Merkow J, Long J, et al. CHEXSTRAY: REAL-TIME MULTI-MODAL DATA CONCORDANCE FOR DRIFT DETECTION IN MEDICAL IMAGING AI. Published online 2022. Accessed November 14, 2022. https://github.com/microsoft/MedImaging-ModelDriftMonitoring

21. Rabanser S, Unnemann SG¨, Lipton ZC. Failing Loudly: An Empirical Study of Methods for Detecting Dataset Shift. Accessed November 14, 2022. https://tensorflow.org/tfx/data_validation/get_started#checking_data_skew_and_drift

22. Liu F, Xia Y, Yang D, Yuille A, Xu D. An Alarm System for Segmentation Algorithm Based on Shape Model.

23. Robinson R, Valindria V V., Bai W, et al. Automated quality control in image segmentation: Application to the UK Biobank cardiovascular magnetic resonance imaging study. *Journal of Cardiovascular Magnetic Resonance*. 2019;21(1):1-14. doi:10.1186/S12968-019-0523-X/FIGURES/7

24. Jin X, Thomas MA, Dise J, et al. Robustness of deep learning segmentation of cardiac substructures in noncontrast computed tomography for breast cancer radiotherapy. *Med Phys*. 2021;48(11):7172-7188. doi:10.1002/MP.15237

25. He K, Zhang X, Ren S, Sun J. Deep residual learning for image recognition. In: *Proceedings of the IEEE Computer Society Conference on Computer Vision and Pattern Recognition*. Vol 2016-December. IEEE Computer Society; 2016:770-778. doi:10.1109/CVPR.2016.90

26. Kingma DP, Welling M. Auto-Encoding Variational Bayes. *2nd International Conference on Learning Representations, ICLR 2014 - Conference Track Proceedings*. Published online December 20, 2013. doi:10.48550/arxiv.1312.6114

27. Zhou D, Yu Z, Xie E, et al. Understanding The Robustness in Vision Transformers. Published online June 28, 2022:27378-27394. Accessed November 29, 2022. https://proceedings.mlr.press/v162/zhou22m.html

28. Zhang C, Zhang M, Zhang S, et al. Delving Deep Into the Generalization of Vision Transformers Under Distribution Shifts. Published online 2022:7277-7286. Accessed November 29, 2022. https://github.com/Phoenix1153/ViT

29. Bai Y, Mei J, Yuille A, Xie C. Are Transformers More Robust Than CNNs? *Adv Neural Inf Process Syst*. 2021;32:26831-26843. doi:10.48550/arxiv.2111.05464

30. Ronneberger O, Fischer P, Brox T. U-net: Convolutional networks for biomedical image segmentation. In: *Lecture Notes in Computer Science (Including Subseries Lecture Notes in Artificial Intelligence and Lecture Notes in Bioinformatics)*. Vol 9351. Springer Verlag; 2015:234-241. doi:10.1007/978-3-319-24574-4_28

31. Vaswani A, Brain G, Shazeer N, et al. *Attention Is All You Need*.

32. Chen C, Dou Q, Chen H, Qin J, Heng PA. Unsupervised Bidirectional Cross-Modality Adaptation via Deeply Synergistic Image and Feature Alignment for Medical Image Segmentation. *IEEE Trans Med Imaging*. 2020;39(7):2494-2505. Accessed May 5, 2021. http://arxiv.org/abs/2002.02255

33. Van Den Oord A, Vinyals O, Kavukcuoglu K. Neural Discrete Representation Learning. *Adv Neural Inf Process Syst*. 2017;2017-December:6307-6316. doi:10.48550/arxiv.1711.00937

34. Chen X, Men K, Chen B, et al. CNN-Based Quality Assurance for Automatic Segmentation of Breast Cancer in Radiotherapy. *Front Oncol*. 2020;10:524. doi:10.3389/FONC.2020.00524/BIBTEX

35. Robinson R, Oktay O, Bai W, et al. Real-time Prediction of Segmentation Quality.

36. Liu Z, Mhlanga JC, Siegel BA, Jha AK. Need for objective task-based evaluation of AI-based segmentation methods for quantitative PET. *ArXiv*. Published online March 1, 2023. Accessed April 13, 2023. /pmc/articles/PMC10002799/

37. Jha AK, Myers KJ, Obuchowski NA, et al. Objective Task-Based Evaluation of Artificial Intelligence-Based Medical Imaging Methods:: Framework, Strategies, and Role of the Physician. *PET Clin*. 2021;16(4):493-511. doi:10.1016/J.CPET.2021.06.013

38. Jha AK, Bradshaw TJ, Buvat I, et al. Nuclear Medicine and Artificial Intelligence: Best Practices for Evaluation (the RELAINCE guidelines). *Journal of Nuclear Medicine*. 2022;63(9):1288-1299. doi:10.2967/JNUMED.121.263239

39. Jha AK, Mena E, Caffo B, et al. Practical no-gold-standard evaluation framework for quantitative imaging methods: application to lesion segmentation in positron emission tomography. *Journal of Medical Imaging*. 2017;4(1):011011. doi:10.1117/1.JMI.4.1.011011